\documentclass[12pt,draftclsnofoot,onecolumn,peerreview]{IEEEtran}
\usepackage{textcomp}
\usepackage{enumerate}
\usepackage{QED}
\usepackage{algorithmic}
\usepackage{mathrsfs}
\usepackage{amsfonts}
\usepackage{amsmath}
\usepackage{amssymb}
\usepackage{cite}      % Written by Donald Arseneau
\renewcommand\citepunct{, }

\newtheorem{proposition}{Proposition}

\makeatletter
\def\Ddots{\mathinner{\mkern1mu\raise\p@
\vbox{\kern7\p@\hbox{.}}\mkern2mu
\raise4\p@\hbox{.}\mkern2mu\raise7\p@\hbox{.}\mkern1mu}}
\makeatother

\renewcommand\thepage{}
\renewcommand{\baselinestretch}{1.75}

\usepackage{graphicx}  % Written by David Carlisle and Sebastian Rahtz
\begin{document}
\title{An Iterative Geometric Mean Decomposition Algorithm for MIMO Communications Systems}
%\author{Chiao-En Chen,~\IEEEmembership{Member,~IEEE}}

\author{Chiao-En Chen, ~\IEEEmembership{Member,~IEEE} and Chia-Hsiang Yang, ~\IEEEmembership{Member,~IEEE} \thanks{Chiao-En Chen is with the Department of Electrical Engineering,
National Chung Cheng University, Chiayi, Taiwan, R.O.C. (e-mail: ieecec@ccu.edu.tw).} \thanks{Chia-Hsiang Yang is with the Department of Electronics Engineering, National Chiao
Tung University, Hsinchu, Taiwan, R.O.C.}}

\maketitle

\begin{abstract}
This correspondence presents an iterative geometric mean decomposition (IGMD) algorithm for multiple-input-multiple-output (MIMO) wireless communications. In contrast to the existing GMD algorithms, the proposed IGMD does not require the explicit computation of the geometric mean of positive singular values of the channel matrix, and hence substantially reduces the required hardware complexity. The proposed IGMD has a regular structure and can be easily adapted to solve problems with different dimensions. We show that the proposed IGMD is guaranteed to converge to the perfect 
GMD under certain sufficient condition. Three different constructions of the proposed algorithm are proposed and compared through computer simulations. Numerical results show that the proposed algorithm quickly attains comparable performance to that of the true GMD within only a few iterations.
\end{abstract}



% Will be ignored for other modes.
\IEEEpeerreviewmaketitle

\section{Introduction}
\PARstart{M}{ultiple}-input-multiple-output (MIMO) communications \cite{Telatar:99:ETT,Foschini:96_BLTJ} have continued to be one of the key technologies of the
next generation wireless systems due to their potential of providing higher data rate and better reliability 
compared to the conventional single-input-single-output (SISO) systems. When the channel state information (CSI)
is available at both the transmitter and receiver, it is well known that the closed-loop gain can be further acquired
by jointly designing the precoder and the equalizer. Among these closed-loop transceiver design schemes, singular-value-decomposition (SVD)-based linear transceiver decomposes the MIMO channel into multiple parallel subchannels and is known to achieve the channel capacity if proper power allocation \cite{Raleigh:98_IEEE_TC} is applied. However, due to the variation of the signal-to-noise-ratio (SNR) in each subchannel, the bit error rate (BER) performance is dominated by the subchannel with the worst SNR. Consequently, without sophisticated bit-allocation schemes, fundamental trade-off between the BER and capacity cannot be avoided in this type of design \cite{Jiang:05_IEEE_TSP_Joint,Jiang:05_IEEE_TSP}.

In addition to the SVD-based linear design, geometric-mean-decomposition (GMD)-based nonlinear transceiver design has also been proposed \cite{Jiang:05_IEEE_TSP_Joint,Zhang:05_IEEE_TIT}. With the help of GMD \cite{Jiang:05_LAA,Zhang:05_IEEE_TIT}, the MIMO channel is decomposed into multiple subchannels with identical SNR, and hence the simple identical bit allocation can be used for all subchannels. It has also been shown that the GMD-based transceiver under the zero-forcing (ZF) constraint asymptotically achieves both the optimal BER and capacity simultaneously at sufficiently high SNR. Due to these nice properties, various extensions and generalizations of GMD-based transceivers have been reported in the literature \cite{Jiang:05_IEEE_TSP,Lin:08_IEEE_TWC,Liu:10_TVT,Weng:11_IEEE_TSP,Liu:12_IEEE_TSP_part1,Liu:12_IEEE_TSP_part2}. 

As the GMD is the core of many advanced MIMO transceiver designs, the associated implementation issues started to draw researchers' attention \cite{Kan:07_ISCS,Chen:13_IEEE_TVT}. In \cite{Kan:07_ISCS}, a scaled geometric mean decomposition algorithm was proposed in order to simplify the detection logic. In \cite{Chen:13_IEEE_TVT}, the authors presented a constant throughput of GMD implementation which also supports hardware sharing between precoding and signal detection modules. In this correspondence, a new implementation issue of the GMD algorithm is addressed. We noticed that all the existing GMD algorithm requires the computation of the geometric mean $\bar{\sigma}$ of all the positive singular values of the matrix to be decomposed in the initialization step. This requires the capability of computing the $K$th root $\sqrt[K]{A}$ of some positive real number $A=\prod_{i=1}^K \sigma_i$, in which $\sigma_i>0$ for all $i=1,\ldots, K$. For special cases where $K=2^L$ with $L$ being some positive integer, it is possible to decompose the computation of $\bar{\sigma}$ into successive geometric mean computations of two numbers,
\begin{align*}
\bar{\sigma}
&=\sqrt{\cdots\sqrt{\sqrt{\sigma_1\sigma_2}\sqrt{\sigma_3\sigma_4}}\cdots\sqrt{\sqrt{\sigma_{K-3}\sigma_{K-2}}\sqrt{\sigma_{K-1}\sigma_K}}},
\end{align*}
where the square root operation can be carried out efficiently using CORDIC-based computing \cite{Meher_IEEE_CAS1}. For $K\neq 2^L$, while finding 
$\sqrt[K]{A}$ can generally be achieved by using Newton's type of $K$th-root algorithm \cite{Atkinson:89_book}, the difficulty lies in the fact that a good initial guess is often required for the algorithm to converge. Another possible way of computing $\sqrt[K]{A}$ is to first transform $A$ into logarithmic domain and then convert it back after divided by $K$. CORDIC-based algorithms can be used to implement algorithmic and exponential functions, but the inherent bounded input range often limits their applications unless extra pre- and post-processing steps are applied. An alternative method is to use lookup tables and/or a piecewise polynomial (including linear) algorithms to realize both the logarithmic and exponential functions. A mass of memory and extra computations are therefore required to ensure the accuracy for such high dynamic-range computations.

In this correspondence, we propose an iterative GMD (IGMD) algorithm based on the technique of successive approximation. The proposed algorithm has a regular structure and is applicable to matrices of any dimension. It does not require explicit computation of the $K$th root, and hence eliminates the hardware difficulties mentioned above. The convergence proof of proposed iterative GMD algorithm is provided, and numerical results show the performance of the proposed algorithm converges to that of the GMD very quickly within a few iterations.

\textit{Notations:} Throughout this paper, matrices and vectors are
set in boldface, with uppercase letters for matrices and lower case
letters for vectors. The superscripts $^\mathrm{T}$, $^\mathrm{H}$ denote the
transpose and conjugate transpose of a matrix, respectively. $\mathrm{diag}\{x_1,\ldots,x_K\}$ denotes the diagonal matrix with diagonal elements $\{x_1,\ldots,x_K\}$.  $[\mathbf{X}]_{p,q}$ and $[\mathbf{X}]_{m:n,p:q}$ denote the $(p,q)$th component and the submatrix formed by the consecutive $m$th to $n$th rows and $p$th to $q$th columns of $\mathbf{X}$, respectively.

\section{Proposed Iterative Geometric Mean Decomposition Algorithm}
\label{sec:proposed_GMD}

The proposed iterative GMD algorithm can be described as follows.  

\textbf{Initialization:}
Given the matrix $\mathbf{H}\in \mathbb{C}^{N\times M}$ of rank $K\leq \min(N,M)$, The algorithm starts with some general orthogonal decomposition of $\mathbf{H}$
\begin{align}
\mathbf{H}=\mathbf{\tilde{U}}\mathbf{\tilde{R}}\mathbf{\tilde{V}}^{\mathrm{H}},
\label{eq:H_URV}
\end{align}
where $\mathbf{\tilde{U}}\in \mathbb{C}^{N\times K}$ and $\mathbf{\tilde{V}}\in \mathbb{C}^{M\times K}$ are both semi-unitary, 
and $\mathbf{\tilde{R}}\in \mathbb{C}^{K\times K}$ is upper-triangular. For general $N$, $M$, and $K$, one can always choose the singular value decomposition for initialization. For special cases where $\mathbf{H}$ is full column rank with $M=K$, other orthogonal decompositions such as the QR decomposition \cite{Golub:96_book} can also be used.

The algorithm then initializes by setting $\mathbf{Q}=\mathbf{\tilde{U}}$, $\mathbf{S}=\mathbf{\tilde{V}}$,  $\mathbf{R}=\mathbf{\tilde{R}}$, and starts with 
iteration index $\ell=1$.

\textbf{Iteration:} In each iteration, the algorithm performs $K-1$ stages of operations as the stage index $k$ ranges from $1$ to $K-1$. 

At stage $k$, the algorithm first computes the singular value decomposition for the $2\times 2$ submatrix of $\mathbf{R}$
\begin{align}
\mathbf{R}_{k:k+1,k:k+1}=\left[\begin{array}{cc} R_{k,k} & R_{k,k+1} \\ 0 & R_{k+1,k+1}   \end{array}\right]=\mathbf{U}^{(k)}_\gamma \boldsymbol{\Sigma}^{(k)}_\gamma \mathbf{V}_\gamma^{(k)\mathrm{H}},
\label{R_tilde_22}
\end{align}  
where the singular matrices $\mathbf{U}^{(k)}_\gamma \in \mathbb{C}^{2\times 2}$ and $\mathbf{V}^{(k)}_\gamma \in \mathbb{C}^{2\times 2}$ are both unitary, and 
$\boldsymbol{\Sigma}^{(k)}_\gamma=\mathrm{diag}\left\{ \sigma^{(k)}_{\gamma,1} , \sigma^{(k)}_{\gamma,2}  \right\}$ is a diagonal matrix with singular values $\sigma^{(k)}_{\gamma,1}$ and $\sigma^{(k)}_{\gamma,2}$. Without loss of generality, we assume $\sigma^{(k)}_{\gamma,2}\leq \sigma^{(k)}_{\gamma,1}$. After the singular values are obtained, carefully designed planar rotations are then applied to obtain an upper triangular matrix with positive diagonal elements $\Omega\left(R_{k,k},R_{k+1,k+1}\right)$ and $R_{k,k}R_{k+1,k+1}/\Omega\left(R_{k,k},R_{k+1,k+1}\right)$, where $\Omega$ is a continuous mapping from $(0,\infty)\times (0,\infty)$ to $(0,\infty)$ with some desired property to be discussed in details shortly. In matrix notations, we then have
\begin{align}
\boldsymbol{\Phi}^{(k)}_\mathrm{L}\boldsymbol{\Sigma}^{(k)}_\gamma\boldsymbol{\Phi}^{(k)}_\mathrm{R}=
\left[\begin{array}{cc} \Omega\left(R_{k,k},R_{k+1,k+1}\right) & \star \\ 0 & \frac{R_{k,k}R_{k+1,k+1}}{\Omega\left(R_{k,k},R_{k+1,k+1}\right)} \end{array}\right].
\label{eq:planar_rotation_v1}
\end{align}
Note that the planar rotations $\boldsymbol{\Phi}^{(k)}_\mathrm{L}$ and $\boldsymbol{\Phi}^{(k)}_\mathrm{R}$ applied in (\ref{eq:planar_rotation_v1}) always exist as long as $\left[\sigma^{(k)}_{\gamma,1},\sigma^{(k)}_{\gamma,2}\right]^\mathrm{T}$ multiplicatively majorizes
$\left[\Omega\left(R_{k,k},R_{k+1,k+1}\right), R_{k,k}R_{k+1,k+1}/\Omega\left(R_{k,k},R_{k+1,k+1}\right)\right]^\mathrm{T}$, or equivalently when
$\sigma^{(k)}_{\gamma,2}\leq\Omega\left(R_{k,k},R_{k+1,k+1}\right)\leq \sigma^{(k)}_{\gamma,1}$ holds \cite{Weyl:49_Proc_NAS,Marshall:91_book}. It is easy to verify that the matrices $\boldsymbol{\Phi}^{(k)}_\mathrm{L}$ and $\boldsymbol{\Phi}^{(k)}_\mathrm{R}$ can be constructed as
\begin{align}
\boldsymbol{\Phi}^{(k)}_{\mathrm{L}}&=\frac{1}{\Omega\left(R_{k,k},R_{k+1,k+1}\right)}\left[\begin{array}{cc}c\sigma^{(k)}_{\gamma,1} & s\sigma^{(k)}_{\gamma,2} \\ -s\sigma^{(k)}_{\gamma,2} & c\sigma^{(k)}_{\gamma,1}\end{array}\right],\label{eq:Phi_L}\\
\boldsymbol{\Phi}^{(k)}_{\mathrm{R}}&=\left[\begin{array}{cc}c & -s \\ s & c\end{array}\right],
\end{align}
where
\begin{align}
c=\sqrt{\frac{\Omega\left(R_{k,k},R_{k+1,k+1}\right)^2-\left(\sigma^{(k)}_{\gamma,2}\right)^2}{\left(\sigma^{(k)}_{\gamma,1}\right)^2-\left(\sigma^{(k)}_{\gamma,2}\right)^2}},\; s=\sqrt{1-c^2}.
\end{align}
Combining the relations in (\ref{R_tilde_22}) and (\ref{eq:planar_rotation_v1}), we then obtain
\begin{align}
&\boldsymbol{\Theta}_\mathrm{L}^{(k)}\mathbf{R}_{k:k+1,k:k+1}\boldsymbol{\Theta}_\mathrm{R}^{(k)}=\left[\begin{array}{cc} \Omega\left(R_{k,k},R_{k+1,k+1}\right) & \star \\ 0 & 
\frac{
R_{k,k} R_{k+1,k+1}}{\Omega\left(R_{k,k},R){k+1,k+1}\right)} \end{array}\right],
\label{eq:transformation}
\end{align}
%\begin{align}
%&\boldsymbol{\Theta}_\mathrm{L}^{(k)}\mathbf{R}_{k:k+1,k:k+1}\boldsymbol{\Theta}_\mathrm{R}^{(k)}\nonumber\\
%=&\left[\begin{array}{cc} \Omega\left(R_{k,k},R_{k+1,k+1}\right) & \star \\ 0 & 
%\frac{
%R_{k,k} R_{k+1,k+1}}{\Omega\left(R_{k,k},R){k+1,k+1}\right)} \end{array}\right],
%\label{eq:transformation}
%\end{align}
where $\boldsymbol{\Theta}^{(k)}_\mathrm{L}=\boldsymbol{\Phi}^{(k)}_\mathrm{L}\mathbf{U}_\gamma^{(k)\mathrm{H}}$, and 
$\boldsymbol{\Theta}^{(k)}_{\mathrm{R}}=\mathbf{V}^{(k)}_\gamma\boldsymbol{\Phi}^{(k)}_\mathrm{R}$. Since $\boldsymbol{\Theta}^{(k)}_\mathrm{L}$ and $\boldsymbol{\Theta}^{(k)}_\mathrm{R}$
are both products of unitary matrices, they are unitary matrices as well.

After $\boldsymbol{\Theta}^{(k)}_{\mathrm{L}}$ and $\boldsymbol{\Theta}^{(k)}_{\mathrm{R}}$ are obtained,  
$\mathbf{G}_\mathrm{L}^{(k)}$, and $\mathbf{G}_\mathrm{R}^{(k)}$ are then constructed
from the identity matrix $\mathbf{I}_{K}$ with the 
submatrix $\left[\mathbf{G}_\mathrm{L}^{(k)}\right]_{k:k+1,k:k+1}$ and $\left[\mathbf{G}_\mathrm{R}^{(k)}\right]_{k:k+1,k:k+1}$ replaced by $\boldsymbol{\Theta}_\mathrm{L}^{(k)}$ and $\boldsymbol{\Theta}_\mathrm{R}^{(k)}$, respectively.
The matrices $\mathbf{R}$, $\mathbf{Q}$, and $\mathbf{S}$ are
then updated as
\begin{align}
\mathbf{R}&=\mathbf{G}^{(k)}_\mathrm{L}\mathbf{R}\mathbf{G}^{(k)}_\mathrm{R}, \label{eq:update_R}\\
\mathbf{Q}&=\mathbf{Q}\mathbf{G}^{(k)\mathrm{T}}_\mathrm{L},\\
\mathbf{S}&=\mathbf{S}\mathbf{G}^{(k)}_\mathrm{R}. \label{eq:update_S}
\end{align}
It is clear that $\mathbf{R}$ remains upper-triangular, and $\mathbf{Q}$ and $\mathbf{S}$ both remain unitary after  (\ref{eq:update_R})-(\ref{eq:update_S}) are performed at the end of each stage. If the stage index $k$ is smaller than $K-1$, the algorithm set $k=k+1$ and performs the procedure (\ref{R_tilde_22})-(\ref{eq:update_S}). Otherwise, the algorithm set the iteration index $\ell=\ell+1$ and start a new iteration unless the prescribed number of iterations is attained.

For the convenience of the subsequent discussion, we denote $\mathbf{Q}^{[\ell]}$, $\mathbf{R}^{[\ell]}$, $\mathbf{S}^{[\ell]}$ as the updated $\mathbf{Q}$, $\mathbf{R}$, $\mathbf{S}$ respectively at the end of $(K-1)$th stage in the $\ell$th iteration. 
Then the following relations hold for the proposed iterative geometric mean decomposition algorithm:
\begin{align}
\mathbf{Q}^{[\ell+1]}=&\mathbf{Q}^{[\ell]}\mathbf{G}^{(1)\mathrm{T}}_\mathrm{L}\ldots\mathbf{G}^{(K-1)\mathrm{T}}_\mathrm{L},\label{eq:Q_recursive}\\
\mathbf{S}^{[\ell+1]}=&\mathbf{S}^{[\ell]}\mathbf{G}^{(1)}_\mathrm{R}\ldots\mathbf{G}^{(K-1)}_\mathrm{R},\\
\mathbf{R}^{[\ell+1]}=&\mathbf{G}^{(K-1)}_\mathrm{L}\cdots\mathbf{G}^{(1)}_\mathrm{L}\mathbf{R}^{[\ell]}\mathbf{G}^{(1)}_\mathrm{R}\ldots\mathbf{G}^{(K-1)}_\mathrm{R}.
\label{eq:R_recursive}
\end{align}
for all $\ell=0,1,\ldots$. Here $\mathbf{Q}^{[0]}$, $\mathbf{S}^{[0]}$, and $\mathbf{R}^{[0]}$ are defined as
$\mathbf{\tilde{U}}$, $\mathbf{\tilde{V}}$, and $\mathbf{\tilde{R}}$, respectively. The planary rotation matrices $\left\{\mathbf{G}^{(k)}_\mathrm{L}\right\}_{k=1}^{K-1}$ and $\left\{\mathbf{G}^{(k)}_\mathrm{R}\right\}_{k=1}^{K-1}$ clearly also depend on the iteration index $\ell$, but the dependency is not denoted explicitly in (\ref{eq:Q_recursive})-(\ref{eq:R_recursive}) for simplicity as long as no confusion results.

In the following section, we show that one can design the mapping $\Omega$ such that 
\begin{align}
\lim_{\ell\rightarrow \infty} \left[\mathbf{R}^{[\ell]}\right]_{k,k}=\bar{\sigma}=\left(\prod_{k=1}^K \left[\mathbf{\tilde{R}}\right]_{k,k}\right)^{1/K}, 
\end{align}
for all $k=1,\ldots, K$. The Geometric Mean Decomposition is therefore obtained when the algorithm converges.

\section{Design of the mapping $\Omega$}
\label{sec:Design_mapping}

For the ease of following discussions, we introduce several new notations. For given $\tau>0$, we define a subset $\mathbb{A}(\tau)\subset \mathbb{R}^K$:
\begin{align*}
\mathbb{A}(\tau)=\left\{\mathbf{x} \in \mathbb{R}^K\left| x_k>0\; \mathrm{for\;all\;}k=1,\ldots,K,\;\prod_{k=1}^{K}x_k=\tau\right.  \right\},
\end{align*}
where $\mathbf{x}=[x_1,\ldots,x_K]^\mathrm{T}$. We also define continuous mappings $T^{(j)}:\;\mathbb{A}(\tau)\rightarrow \mathbb{A}(\tau),\; j=1,\ldots,K-1$, given by
\begin{align}
T^{(j)}\left(\left[\begin{array}{c} \mathbf{x}_{1:j-1} \\ x_{j}\\ x_{j+1}\\ \mathbf{x}_{j+2:K} \end{array}\right]  \right)&=
\left[\begin{array}{c} \mathbf{x}_{1:j-1} \\ \Omega\left(x_j,x_{j+1}\right)\\ \frac{x_j x_{j+1}}{\Omega\left(x_j,x_{j+1}\right)} \\ \mathbf{x}_{j+2:K} \end{array}\right]. \label{eq:Tj}
\end{align}
If we denote the vector on the main diagonal of $\mathbf{R}^{[\ell]}$ as $\mathbf{r}^{[\ell]}$, then the diagonal vectors of $\mathbf{R}^{[\ell+1]}$ and $\mathbf{R}^{[\ell]}$ can be related from (\ref{eq:update_R}) and (\ref{eq:R_recursive}) using the new notations
\begin{align}
\mathbf{r}^{[\ell+1]}
&=T^{(K-1)}\left( \ldots\left(T^{(2)}\left(T^{(1)}\left(\mathbf{r}^{[\ell]}\right)\right)\right)\ldots\right)\label{eq:definition_T}\\
&=T\left(\mathbf{r}^{[\ell]}\right)=T^{\ell+1}\left(\mathbf{r}^{[0]}\right),
\end{align}
where 
$T(\mathbf{x})=\left(T^{(K-1)}\circ T^{(K-2)}\circ\cdots\circ T^{(2)} \circ T^{(1)}\right)(\mathbf{x})$, $T^{\ell+1}(\mathbf{x})$ is the $(\ell+1)$-fold repeated composition of the mapping $T(\mathbf{x})$, and $\mathbf{r}^{[0]}=\mathrm{diag}\left\{\mathbf{\tilde{R}}\right\}$.

With the aforementioned notations, we now present the main results for the $\Omega$ design.
\begin{proposition}
$\lim_{\ell\rightarrow \infty} \mathbf{r}^{[\ell]}=\bar{\sigma}\mathbf{1}$ if 
the mapping $\Omega:(0,\infty)\times (0,\infty)\rightarrow (0,\infty)$ satisfies the following property
\begin{align}
\Omega(z_1,z_2)+\frac{z_1z_2}{\Omega(z_1,z_2)}\leq z_1+z_2, 
\label{eq:condition_omega}
\end{align}
for all $z_1,z_2>0$, and the equality holds if and only if $z_1=z_2$.
\end{proposition}
\begin{Proof}
Given 
$\mathbf{r}^{[0]}=\left[r^{[0]}_1,\ldots,r^{[0]}_K\right]=\mathrm{diag}\{\mathbf{\tilde{R}}\}$, 
we let $\tau=\prod_{k=1}^{K}r^{[0]}_k$, and consider the function $F:\mathbb{A}(\tau)\rightarrow (0,\infty)$, $F(\mathbf{x})=\sum_{k=1}^K x_k$. From the Arithmetic-Geometric inequality, 
\begin{align}
F(\mathbf{x})=\sum_{k=1}^K x_k\geq K\left(\prod_{k=1}^K x_k\right)^{1/K}=K\tau^{1/K}=K\bar{\sigma},
\end{align}
it is clear that $F(\mathbf{x})$ attains its absolute minimum in $\mathbb{A}(\tau)$ at $\mathbf{x}=\bar{\sigma}\mathbf{1}$. 
From the definition of $T(\mathbf{x})$ and (\ref{eq:condition_omega}), it is clear that
\begin{align}
F\left(T^{(j)}\left(\mathbf{x}\right)\right)&=\sum_{k=1}^{j-1}x_k+\sum_{k=j+2}^{K}x_k+\Omega(x_j,x_{j+1})+\frac{x_jx_{j+1}}{\Omega(x_j,x_{j+1})}\nonumber\\
&\leq \sum_{k=1}^{K}x_k=F(\mathbf{x}) \label{eq:FTj_ineq},
\end{align}
for all $j=1,\ldots, K-1$, and the equality holds if and only if $x_j=x_{j+1}$. 
It follows readily that $T(\mathbf{x})$, which is a composite mapping of $T^{(1)},\ldots,T^{(K)}$, 
satisfies 
\begin{align}
F\left(T(\mathbf{x})\right)\leq F(\mathbf{x}),
\end{align}
with the equality holds if and only if $x_1=x_2=\ldots=x_K$. Consequently, $y^{[\ell-1]}=F(T(\mathbf{r}^{[\ell-1]}))$ is a monotonically decreasing sequence in $(0,\infty)$, and hence is guaranteed to converge to the
greatest lower bound $K\bar{\sigma}$ \cite{Rudin:76_book}. Furthermore, since $F$ is continuous, we have 
\begin{align}
\lim_{\ell\rightarrow \infty}F\left(T\left(\mathbf{r}^{[\ell-1]}\right)\right)=F\left(\lim_{\ell\rightarrow \infty}T\left(\mathbf{r}^{[\ell-1]}\right)\right)=K\bar{\sigma},
\end{align}
which is attained when $\lim_{\ell \rightarrow \infty}T\left(\mathbf{r}^{[\ell-1]}\right)=\bar{\sigma}\mathbf{1}$.
As a result, we have  $\lim_{\ell \rightarrow \infty}\mathbf{r}^{[\ell]}=\bar{\sigma}\mathbf{1}$, which completes the proof.
\end{Proof}

There exists potentially many functions that satisfy condition (\ref{eq:condition_omega}). The geometric mean $\Omega_{\mathrm{GM}}(z_1,z_2)=\sqrt{z_1z_2}$ clearly satisfies (\ref{eq:condition_omega}) as
\begin{align}
\Omega_{\mathrm{GM}}(z_1,z_2)+\frac{z_1z_2}{\Omega_{\mathrm{GM}}(z_1,z_2)}=2\sqrt{z_1z_2}\leq z_1+z_2, 
\end{align}
always holds due to the AM-GM inequality, and equality holds if and only if $z_1=z_2$. In addition to $\Omega_{\mathrm{GM}}(z_1,z_2)$, the arithmetic mean $\Omega_{\mathrm{AM}}(z_1,z_2)=(z_1+z_2)/2$ is another choice that also satisfies (\ref{eq:condition_omega}). This can be observed by squaring both sides of the AM-GM inequality 
\begin{align}
&\;4z_1z_2\leq (z_1+z_2)^2 \nonumber\\
\Leftrightarrow&\; (z_1+z_2)^2+4z_1z_2\leq 2(z_1+z_2)^2\nonumber\\
\Leftrightarrow&\; \frac{z_1+z_2}{2}+\frac{2z_1 z_2}{z_1+z_2}\leq z_1+z_2 \label{eq:AM_GM_square}
\end{align}
As a result, $\Omega_{\mathrm{AM}}(z_1,z_2)+\frac{z_1 z_2}{\Omega_{\mathrm{AM}}(z_1,z_2)}\leq z_1+z_2$, and the equality holds if and only if $z_1=z_2$. Note that $\frac{z_1 z_2}{\Omega_{\mathrm{AM}}(z_1,z_2)}$ is simply the harmonic mean 
function $\Omega_{\mathrm{HM}}(z_1,z_2)=2z_1z_2/(z_1+z_2)$ while $\Omega_{\mathrm{AM}}(z_1,z_2)=z_1 z_2/\Omega_{\mathrm{HM}}(z_1,z_2)$,
it is clear that $\Omega_{\mathrm{HM}}(z_1,z_2)$ also satisfies (\ref{eq:condition_omega}) from the same relation we obtained in (\ref{eq:AM_GM_square}). 

Based on $\Omega_{\mathrm{AM}}$, $\Omega_{\mathrm{GM}}$, and $\Omega_{\mathrm{HM}}$, we can then construct our iterative GMD algorithms: 
IGMD-AM, IGMD-GM, and IGMD-HM, respectively. Since these mappings are highly nonlinear, it is very challenging to compare the convergence speed of the proposed algorithm in these three constructions. In fact, the convergence behaviour not only depends on the topological property of the mapping but also depends on how the algorithms are initialized. From the implementation point of view, IGMD-AM and IGMD-HM may have some advantages over the 
IGMD-GM as only basic arithmetic operations are required in computing $\Omega_{\mathrm{AM}}(z_1,z_2)$ and $\Omega_{\mathrm{HM}}(z_1,z_2)$ rather than the square root operations required in computing $\Omega_{\mathrm{GM}}(z_1,z_2)$. 

\section{Simulation Results}

In this section, we present some simulation results of the proposed IGMD algorithm under three different constructions: IGMD-AM, IGMD-GM, and IGMD-HM. Throughout the simulation, we assume standard $K\times K$ i.i.d. Rayleigh fading channel in which every element in the channel matrix $\mathbf{H}$ is modelled as zero-mean circularly symmetric complex Gaussian random variable with unit variance. To highlight the applicability of the proposed algorithm in the challenging large $K\neq 2^L$ case, we choose $K=7$ in the simulation. Each simulation point in the figure is averaged over $10^4$ channel realizations.

\begin{figure}[thb]
\centering
\includegraphics[width=0.9\textwidth]{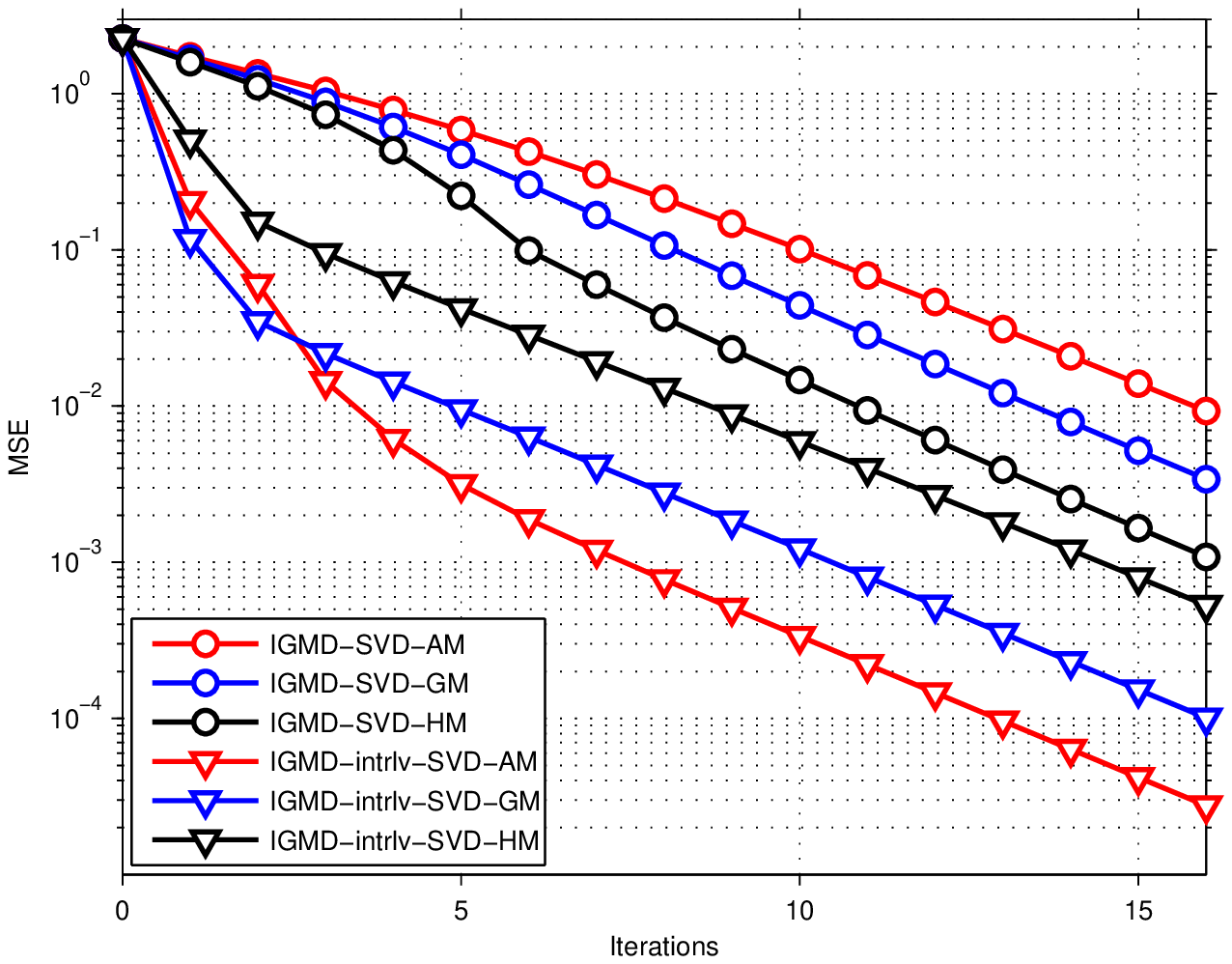}
\caption{MSE comparison of the diagonal elements of $\mathbf{R}$ under proposed Iterative GMD using SVD and interleaved-SVD as initialization.}
\label{fig:convergence_MSE_SVD}
\end{figure}

\begin{figure}[thb]
\centering
\includegraphics[width=0.9\textwidth]{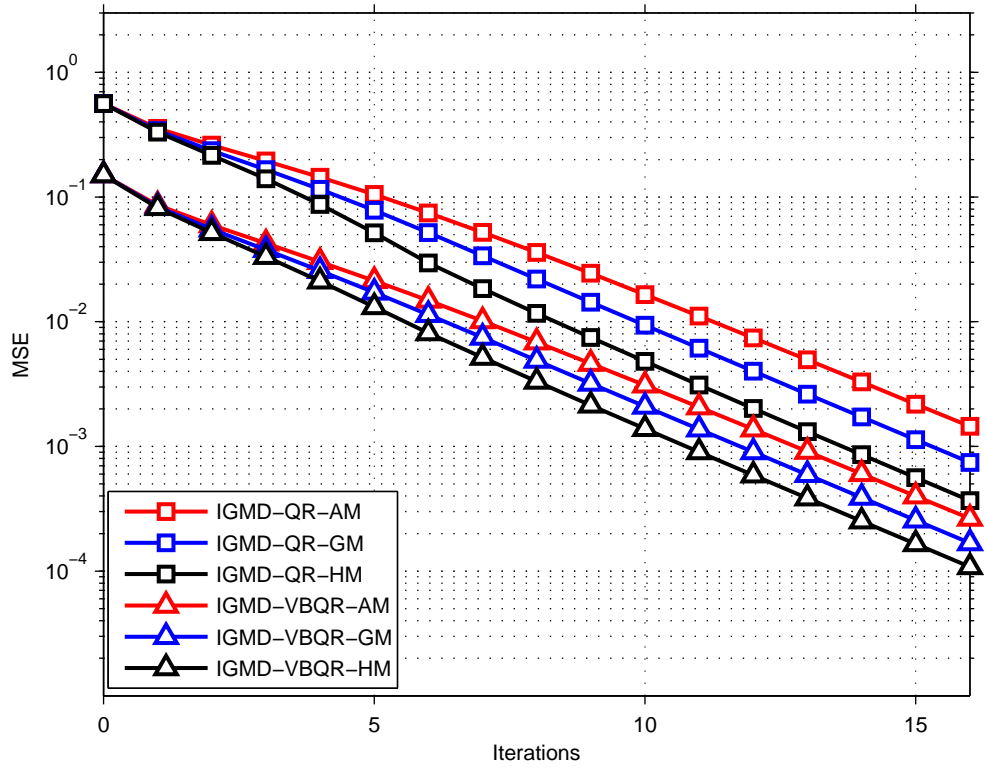}
\caption{MSE comparison of the diagonal elements of $\mathbf{R}$ under proposed Iterative GMD using QR and VB-QR as initialization.}
\label{fig:convergence_MSE_QR}
\end{figure}

Fig. \ref{fig:convergence_MSE_SVD} and Fig. \ref{fig:convergence_MSE_QR} show the mean-square-error (MSE) convergence behaviour of the diagonal elements of $\mathbf{R}$ using SVD and QR as initializations, respectively. For SVD initialization, we propose an alternative interleaved-SVD (intrlv-SVD), defined as
\begin{align}
\mathbf{H}=\mathbf{U}\boldsymbol{\Sigma}\mathbf{V}^{\mathrm{H}}\in \mathbb{C}^{7\times 7},
\end{align}
where $\boldsymbol{\Sigma}=\mathrm{diag}\left\{[\mathrm{\sigma}_1,\mathrm{\sigma}_7,\mathrm{\sigma}_2,\mathrm{\sigma}_6,\mathrm{\sigma}_3,\mathrm{\sigma}_5,\mathrm{\sigma}_4] \right\}$ and $\sigma_1\geq \sigma_{2}\geq \cdots \geq \sigma_7$ to enable more efficient averaging in each stage. For QR initialization, the QR factorization with VBLAST ordering (VBQR) \cite{Wolniansky:98_URSI,Golden:99_EL} is also proposed  to speed up convergence. As the diagonal elements of $\mathbf{\tilde{R}}$ in VBQR generally has less spread than those in QR, VBQR initialization provides a smaller MSE when used in the initialization as shown in Fig. \ref{fig:convergence_MSE_QR}. 

Due to the highly nonlinear nature of the corresponding mappings $T_{\mathrm{AM}}(\cdot)$, $T_{\mathrm{GM}}(\cdot)$, and $T_{\mathrm{HM}}(\cdot)$, it is very difficult to analytically compare the convergence behaviour of the three IGMD constructions. Hence we resort to numerical simulations and leave the more challenging theoretical analysis to our future work. The simulation results in Fig. \ref{fig:convergence_MSE_SVD} and Fig. \ref{fig:convergence_MSE_QR} show that the IGMD-HM has the fastest convergence rate, followed by the IGMD-GM, and the IGMD-AM when QR, VBQR, and SVD are used as initialization. On the other hand, when intrlv-SVD are used, the IGMD-AM has the fastest convergence rate, followed by the IGMD-GM, and finally the IGMD-HM. 

%\begin{figure}[thb]
%\centering
%\includegraphics[width=0.48\textwidth]{./matlab/MSE_SVD}
%\caption{MSE comparison of the diagonal elements of $\mathbf{R}$ under proposed Iterative GMD using SVD and interleaved-SVD as initialization.}
%\label{fig:convergence_MSE_SVD}
%\end{figure}
%\begin{figure}[thb]
%\centering
%\includegraphics[width=0.48\textwidth]{./matlab/MSE_QR}
%\caption{MSE comparison of the diagonal elements of $\mathbf{R}$ under proposed Iterative GMD using QR and VB-QR as initialization.}
%\label{fig:convergence_MSE_QR}
%\end{figure}

\begin{figure}[thb]
\centering
\includegraphics[width=0.9\textwidth]{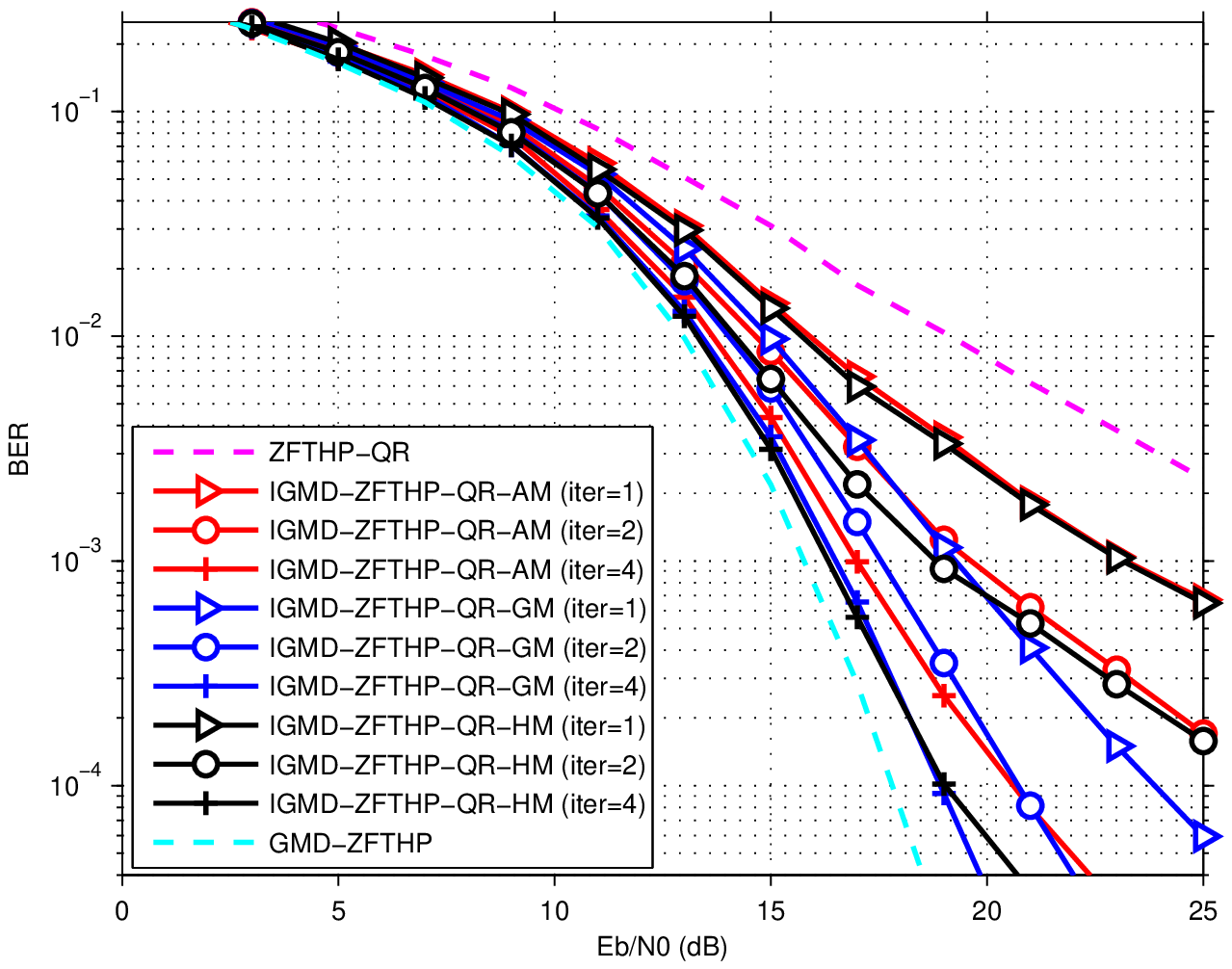}
\caption{BER performance of the proposed Iterative GMD algorithm in a $7\times 7$ MIMO ZF-THP system using QR as initialization.}
\label{fig:convergence_BER_QR}
\end{figure}

\begin{figure}[thb]
\centering
\includegraphics[width=0.9\textwidth]{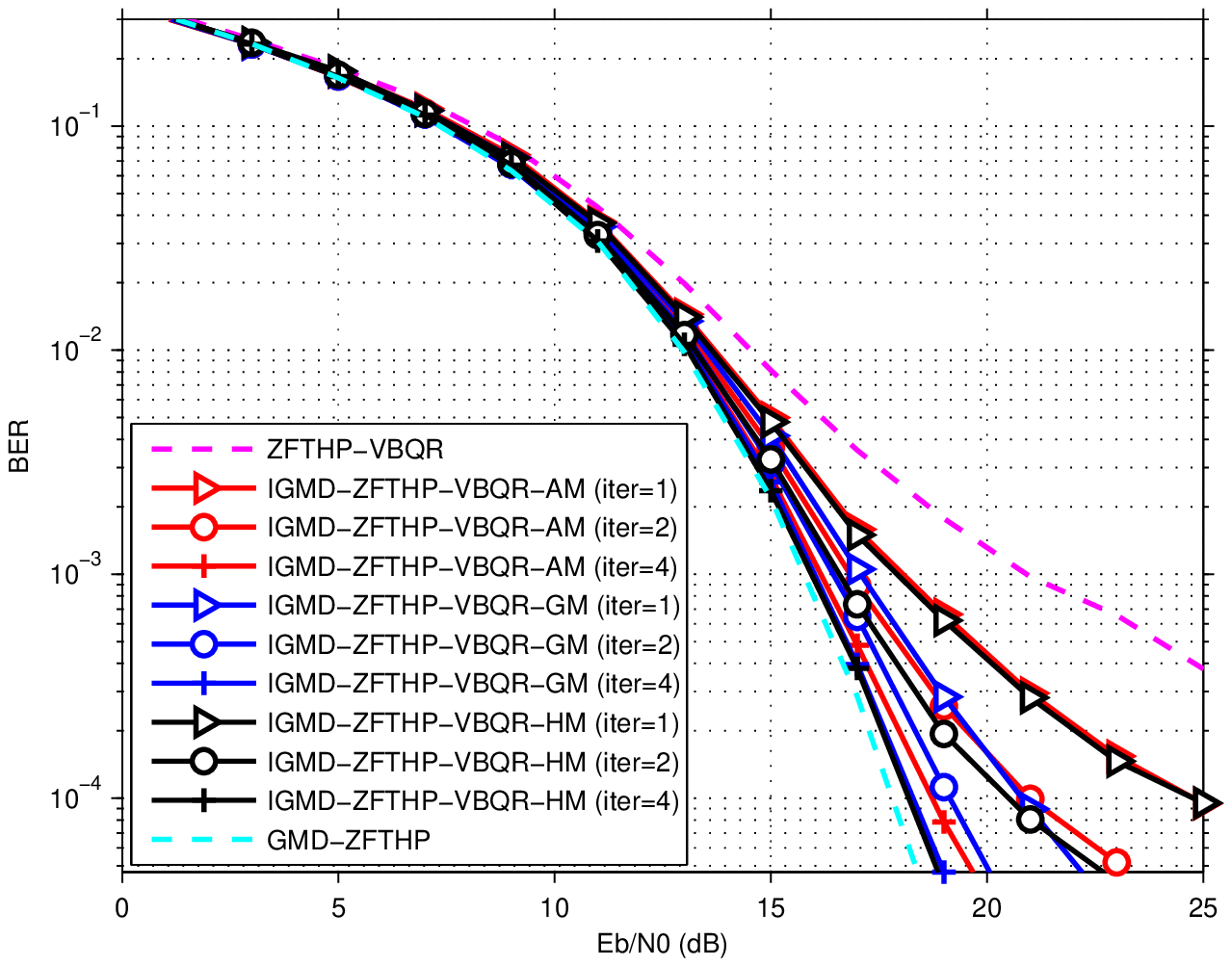}
\caption{BER performance of the proposed Iterative GMD algorithm in a $7\times 7$ MIMO ZF-THP system using VB-QR as initialization.}
\label{fig:convergence_BER_VBQR}
\end{figure}

In the second simulation setting, we investigate the convergence behaviour of the proposed IGMD on the error rate performance of a GMD-based MIMO system. A $7\times 7$ GMD-based zero-forcing Tomlinson-Harashima precoded (ZFTHP) MIMO system \cite{Jiang:05_IEEE_TSP_Joint} using $16$-quadrature amplitude modulation is simulated. Fig. \ref{fig:convergence_BER_QR} and Fig. \ref{fig:convergence_BER_VBQR} show the error rate of the proposed IGMD using QR and  VBQR, respectively. By comparing the BER of ZFTHP-QR in Fig. \ref{fig:convergence_BER_QR} and ZFTHP-VBQR in Fig. \ref{fig:convergence_BER_VBQR}, it is clear that the VBQR provides a better initialization for the proposed IGMD, and results in faster convergence. At the $1$st iteration, the error rate performance of IGMD-AM and IGMD-HM appears to be similar. However, for iteration number greater than $1$, the IGMD-GM and IGMD-HM both outperform the IGMD-AM and perform very close to the optimal GMD after $4$ iterations.

%\begin{figure}[thb]
%\centering
%\includegraphics[width=0.48\textwidth]{./matlab/BER_QR}
%\caption{BER performance of the proposed Iterative GMD algorithm in a $7\times 7$ MIMO ZF-THP system using QR as initialization.}
%\label{fig:convergence_BER_QR}
%\end{figure}
%\begin{figure}[thb]
%\centering
%\includegraphics[width=0.48\textwidth]{./matlab/BER_VBQR}
%\caption{BER performance of the proposed Iterative GMD algorithm in a $7\times 7$ MIMO ZF-THP system using VB-QR as initialization.}
%\label{fig:convergence_BER_VBQR}
%\end{figure}

\begin{figure}[thb]
\centering
\includegraphics[width=0.9\textwidth]{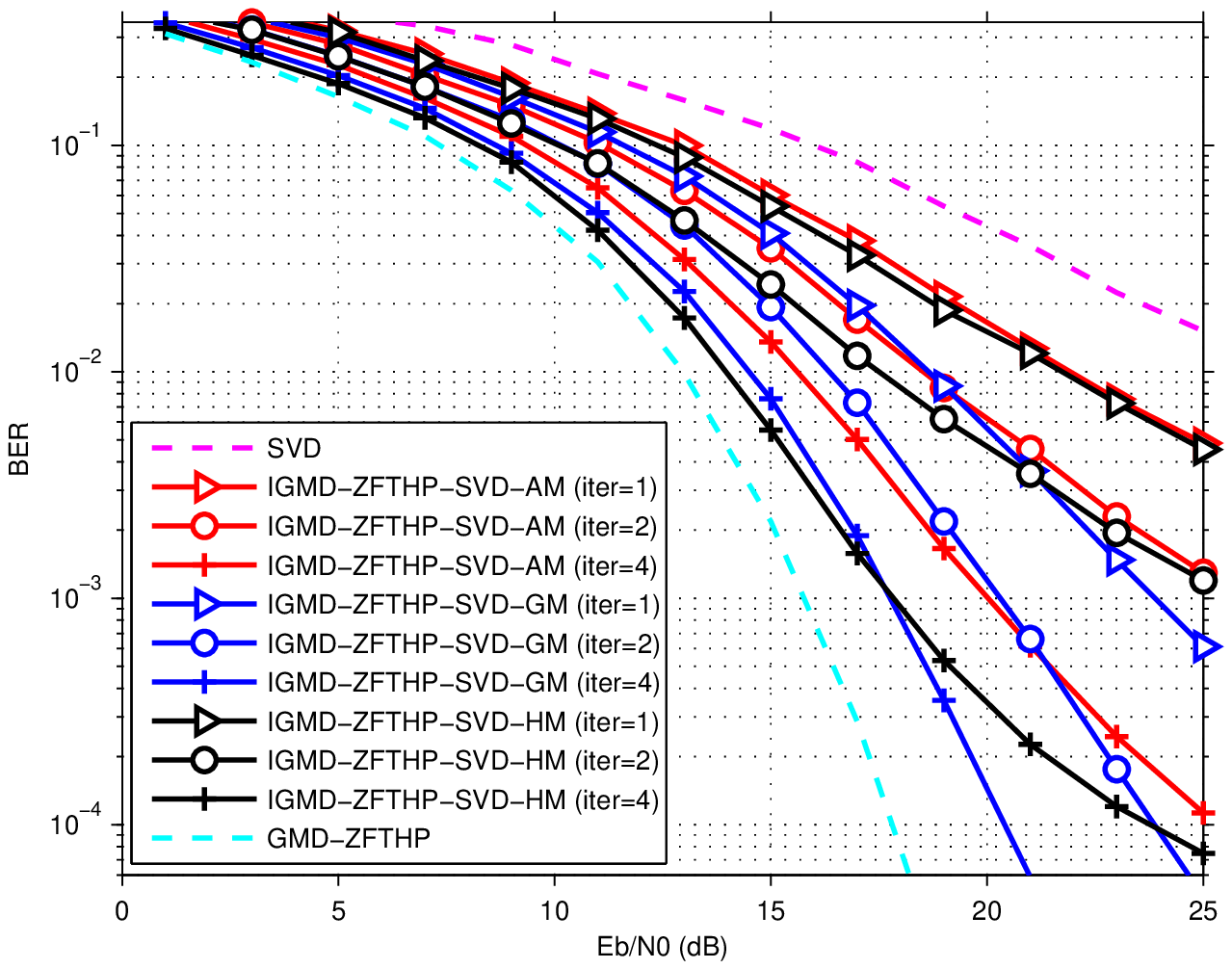}
\caption{BER performance of the proposed Iterative GMD algorithm in a $7\times 7$ MIMO ZF-THP system using SVD as initialization.}
\label{fig:convergence_BER_SVD}
\end{figure}

\begin{figure}[thb]
\centering
\includegraphics[width=0.9\textwidth]{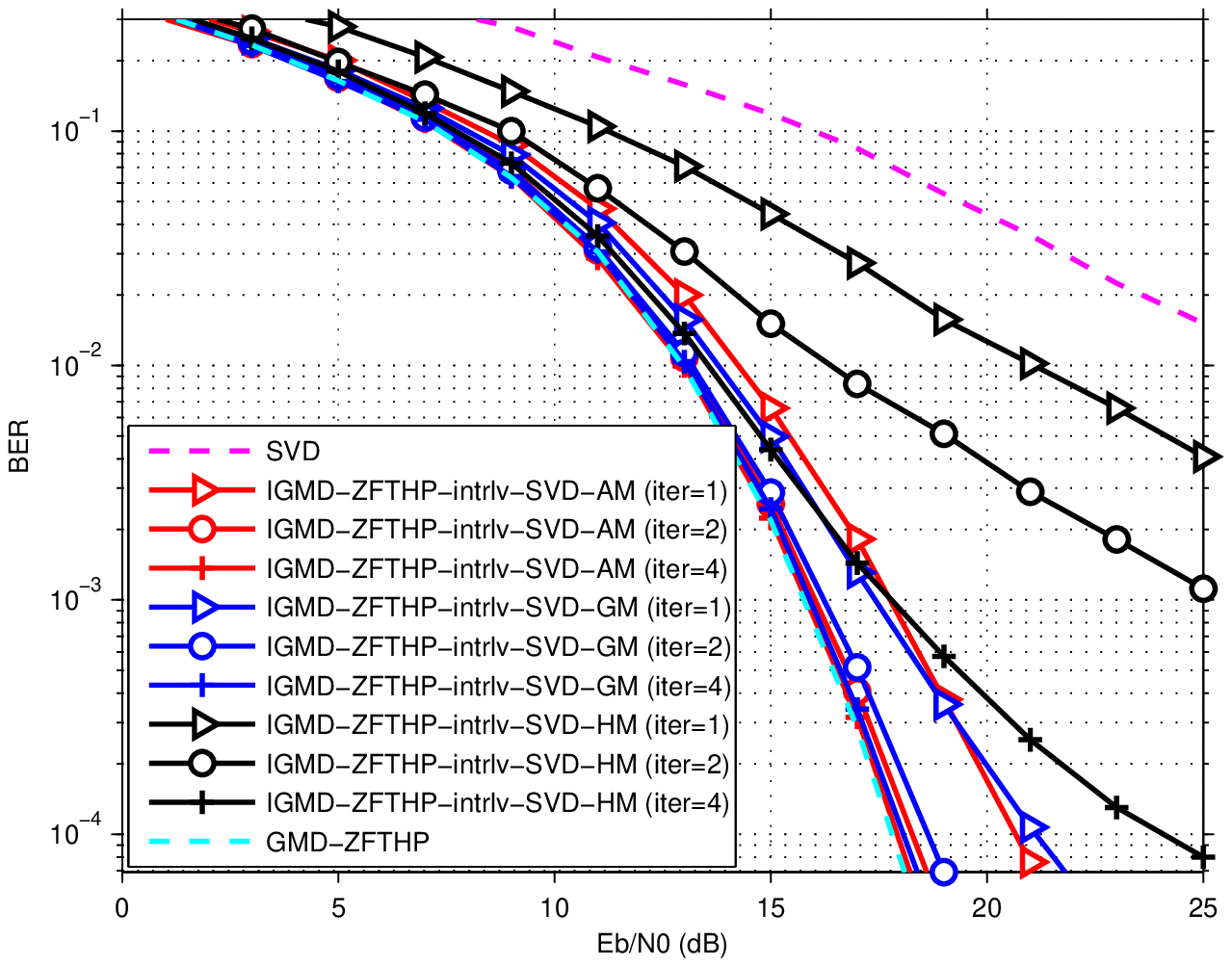}
\caption{BER performance of the proposed Iterative GMD algorithm in a $7\times 7$ MIMO ZF-THP system using interleaved-SVD as initialization.}
\label{fig:convergence_BER_intrlvSVD}
\end{figure}

Fig. \ref{fig:convergence_BER_SVD} and Fig. \ref{fig:convergence_BER_intrlvSVD} show the error rate of the proposed IGMD using regular SVD and interleaved SVD, respectively. In medium to low SNR region, the IGMD-HM shows comparable or even better performance than the IGMD-GM, while for sufficiently high SNR, the IGMD-GM performs the best, followed by the IGMD-HM, and the IGMD-AM. On the contrary, the proposed IGMD using interleaved SVD as initialization shows very different convergence behaviour. The IGMD-HM with interleaved SVD performs much worse compared to the IGMD-AM and IGMD-GM. For most SNR region of practical interests in this setting, the IGMD-AM is comparable to the IGMD-GM for iteration number greater than $1$. From Fig. \ref{fig:convergence_BER_VBQR} and Fig \ref{fig:convergence_BER_intrlvSVD},
it is observed that the proposed IGMD-intrlv-SVD-GM and IGMD-intrlv-SVD-AM achieve even better error rate than the IGMD-VBQR-GM and IGMD-VBQR-HM after $4$ iterations.

%\begin{figure}[thb]
%\centering
%\includegraphics[width=0.48\textwidth]{./matlab/BER_SVD}
%\caption{BER performance of the proposed Iterative GMD algorithm in a $7\times 7$ MIMO ZF-THP system using SVD as initialization.}
%\label{fig:convergence_BER_SVD}
%\end{figure}
%\begin{figure}[thb]
%\centering
%\includegraphics[width=0.48\textwidth]{./matlab/BER_intrlvSVD}
%\caption{BER performance of the proposed Iterative GMD algorithm in a $7\times 7$ MIMO ZF-THP system using interleaved-SVD as initialization.}
%\label{fig:convergence_BER_intrlvSVD}
%\end{figure}

\section{Conclusion}
\label{sec:conclusions}
An iterative geometric mean decomposition algorithm for MIMO communications is proposed. The proposed algorithm has a regular structure and can be easily adapted to accommodate problems of different dimensions. Through iteratively updating the constituents matrices, the algorithm is able to converge to the true GMD without performing the $K$th root computation. The convergence of the algorithm under certain sufficient condition is proved analytically and verified through computer simulations.
 
%In this paper, a new class of Tomlinson-Harashima precoding schemes with blockwise-lattice-reduction has been proposed. In contrast to the conventional THPs with full lattice-reduction, the proposed BLR-aided THPs approximately decouple the downlink channels into multiple subchannels before applying the lattice reduction and hence result in much lower computational complexity. Among the proposed BLR-aided precoders, the proposed BLR-SLNR-THP (LP) designs have parallel THP structures and hence enjoy shorter precoding latency. The precoder designs using SSLNR preprocessing filters, on the other hand, require higher precoding latency but provide better performance than their SLNR counterparts. A computationally efficient inter-cluster ordering algorithm which performs very close to the optimal ordering with exhaustive search has been derived. With the assistance of the proposed inter-cluster ordering algorithm, the BLR-subOSSLNR-MMSE-THP achieves an error rate performance very close to the ultimate LR-MMSE-THP with lower computational complexity. These facts show that the proposed BLR-aided THP designs stand out as useful alternatives in balancing the trade-offs between complexity and performance for MU-MIMO downlink communications in the presence of clusters of correlated users.

\section*{Acknowledgment}
\addcontentsline{toc}{section}{Acknowledgment}
The authors would like to thank Prof. Chiu-Chu Melissa Liu for the helpful discussion.

\bibliographystyle{IEEEtran}
%\bibliography{D:/Assistant_Professor/[Research]/working_paper/MIMO_references}

\newpage

\newpage

\newpage

\newpage

\newpage

\newpage

\end{document}